\documentclass[proceedings, preprint]{rmaa}

\usepackage{paralist}

\usepackage{psfrag,color}

\newcommand{\pasa}{PASA}

\SetYear{2023}
\SetConfTitle{Stellar Feedback in the ISM}

\title{Conference Summary: Stellar Feedback  in the ISM: Celebrating
  the Life and Work of You-Hua Chu
} 

\author{
  Mordecai-Mark Mac Low\altaffilmark{1} 
}

\altaffiltext{1}{Department of Astrophysics, American Museum of
  Natural History, 200 Central Park West, New York, NY 10024, USA (mordecai@amnh.org).}

\shortauthor{Mac Low}
\shorttitle{Conference Summary}

\listofauthors{M.-M. Mac Low}
\indexauthor{Mac Low, M.-M.}

\abstract{In this conference summary I first provide some historical
  notes on my own collaboration with You-Hua Chu and on the discovery
  of X-rays from superbubbles by Margarita Rosado S.  I then
  considere the central subject of the conference, stellar feedback,
  and how interactions between stellar winds and the interstellar
  medium can limit or enhance the effects of feedback compared to
  models including only supernova explosions.  Finally, I review
  results in other areas covered by the conference, including planet and
  formation, nova and supernova remnants, different topics in stellar
  evolution and interaction with the interstellar medium, star
  clusters, observational surveys, and observational and numerical techniques.
  }

\resumen{ }

\addkeyword{H~II regions}
\addkeyword{ISM: Jets and outflows}
\addkeyword{Stars: Pre-main sequence}
\addkeyword{Stars: Mass loss}

\begin{document}
\maketitle

\section{Historical}

\subsection{Personal}
I first met You-Hua Chu while I was a graduate student visiting Michael Norman at the National
Center for Supercomputing Applications (NCSA) at the University of
Illinois in 1987. I listened in to You-Hua and Mike in his office as
they considered the consequences of a
blast wave running through a clumpy interstellar medium
\citep{norman1988}.  
I explained to her that I was working on a thesis on
the theory of superbubbles.  Shortly afterwards, she emailed a
series of cogent questions about X-ray emission from superbubbles,
which my thesis advisor Dick McCray strongly suggested I engage with.
That was excellent advice that led to our description of X-ray
emission from superbubbles in the Large Milky  (Oey and others' suggested decolonial
renaming from Magellanic)  Cloud \citep[LMC;][]{chu1990} using the {\em Einstein} data taken
by Rosado and described in the next section.  Our collaboration was a major factor in my early
career, leading to seven papers together.

We also collaborated in advising students, all of whom have
contributed to this conference.  The first was Guillermo
Garc\'{\i}a-Segura, who after an initial observational paper, turned
out to be a theorist whose work on the structure of bubbles in
time-varying stellar winds has stood the test of time.  The second was
Sean Points, who was most certainly not a theorist, but has had a
distinguished career supporting CTIO and surveying the Magellanic
Clouds.  Finally came Chao-Chin Yang, who led our demonstration that
Toomre gravitational instability of stars and gas can explain the
locations of star formation in the LMC \citep{yang2007}.  He then
joined my group and moved from working on galactic disks to working on
protostellar disks and planet formation, where he continues to have a
substantial impact.

\subsection{X-ray Observations of Superbubbles}

Margarita Rosado Sol\'{\i}s contributed a history of how the first X-ray
observations of superbubbles came to be, which I include here.

\begin{quotation}
Most of you do not know but I have the honor of having contributed
with my small grain of sand to the discovery of X-ray emission from
superbubbles. I was a PhD student when my observing proposal with the
Einstein Satellite of deep exposures of several superbubbles in the
LMC including the superbubbles N70 and N185 was
accepted. Indeed, at that time I have just measured the high expansion
velocities of N7 and N185 by means of Fabry-Perot interferometry
(about 70 km/s) and I have computed the X-ray luminosities submitting
a proposal to the Einstein Observatory together with my adviser Guy
Monnet. The Rosado \& Monnet proposal was accepted by Einstein
Observatory board and the observations carried out giving the result
of the detection, for the first time, of X-ray emission from the
superbubbles N70 and N185, among other superbubbles.

We submitted an article reporting those successful results that was rejected by an anonymous referee that argued that it was only noise (at that time the Einstein Observatory instrumental function was concealed for alien users as me, so that it was really hard to answer to the aggressive referee). Years after, our observations were used by You-Hua Chu and Mordecai Mac Low showing that indeed there was X-ray emission from those superbubbles. In fact, bubbles and superbubbles were unexpected objects that successfully emit in X-rays besides the binary compact X-ray sources. Thanks to that observing proposal the X-ray emission from superbubbles started to be studied from the X-ray observatories.
  \end{quotation}

  \section{Stellar Feedback}
Stellar feedback is required to understand galaxy evolution, as
Pittard summarized in his talk \citep[i.e.][]{dalla_vecchia2008}.
Without including effective stellar feedback, galaxy models form
objects far smaller and denser than observed.  In this
section I summarize how our understanding of this process was advanced
during the conference.

The initial focus of stellar feedback modeling was momentum and energy
transfer from supernovae (SNe).  The momentum injection from SNe in a
uniform medium is well understood \citep[e.g.][]{pittard2019}.  However,
the consequences of an inhomogeneous medium remain controversial, with
\citet{martizzi2015} and \citet{zhang2019} 
finding a 30\% reduction in momentum injection, while \citet{kim2015}
and \citet{walch2015} 
find no reduction.  Numerical algorithms for adding SN feedback
have tended to fail at the numerical resolutions practical for whole
galaxy or cosmological models.  Only recently have examples been described of
apparently resolution-independent algorithms such as FIRE-2
\citep{hopkins2018}.  

However, the story is likely to be more complicated than that for
several reasons.  First, Pittard noted that it is likely that not all
massive stars explode as SNe.  \citet{smartt2015} found that no star
with an initial mass exceeding 20~M$_\odot$ has been observed to
explode. Models by \citet{sukhbold2016} indeed suggest that direct
collapse dominates the outcomes for stars greater than that mass, and
even isolated masses down to as low as 15~M$_\odot$.  Oey noted that
several groups have found that low-metallicity stars have a lower
threshold for direct collapse \citep{heger2003, zhang2008,
oconnor2011, sukhbold2016}.  Second, Oey also noted that there can be
a several megayear delay before SNe begin.  At high densities,
neglecting other forms of stellar feedback such as
stellar winds can lead to dramatically higher star formation
efficiency (SFE).  However, stellar
winds can be an order of magnitude weaker for substantially
subsolar luminosities \citep{jecmen2023}. 

A further puzzle is that observed stellar wind bubbles often appear to
expand too slowly and have too little X-ray emission compared to what would be expected from the
\citet{weaver1977} dynamical model. Chu gave an example from
\citet{naze2001}, showing a 15~pc bubble expanding at only
15--20~km~s$^{-1}$. Oey reviewed the idea that catastrophic cooling of
the hot shocked wind region in the interior can shift the solution
from the energy conserving solution
\citep{pikelner1968,avedisova1972, castor1975,dyson1975,weaver1977} to the
momentum-conserving solution \citep{steigman1975}.  This likely
happens in superbubbles as well.  The example of N79 in the LMC was
described by Rodriguez, presenting results in preparation by Webb \&
Rodriguez.  They find that the diffuse X-ray emission from this super
star cluster is an order of magnitude below that predicted by
\citet{weaver1977} or \citet{chu1990}. Further examples of this
phenomenon given by Oey include
\citet{saken1992,brown1995,oey1996,oey1998,cooper2004,smith2005} and
\citet{oey2009}. Pittard summarized models of efficiently cooling
bubbles, primarily due to turbulent mixing at the interface between
hot and cold gas \citep{rogers2013,geen2015,haid2018,lancaster2021}.
An analytical model of a leaky bubble is an alternative
\citep{harper-clark2009}, although that presupposes a lower-density
region to leak into, which isn't necessarily available.

Oey argued that at low metallicity, super star clusters fail to
effectively drive winds at early times because of the reduced stellar
wind strengths expected.  This leads to high-density gas being
retained near the clusters \citep{jecmen2023}, catastrophic cooling
\citep{silich2004,wuensch2007}, and thus insufficient time to launch a
superwind \citep{danehkar2021}.  Feedback during this period is then
radiation dominated \citep{freyer2003,krumholz2009,komarova2021},
leading to higher SFE \citep{krause2012,silich2018}, greater gas
clumpiness allowing Lyman continuum to escape \citep{jaskot2019}, and
smaller superbubbles.  The discovery by
a group including You-Hua of diffuse nebular \ion{C}{4} 
emission around the slowly expanding superbubble Mrk 71 supports this
scenario \citep{oey2023}.  Even at solar metallicity, models by
\citet{polak2023} find that centrally-concentrated gas clouds with
masses approaching $10^6$~M$_\odot$ have high SFE and do not
effectively drive superwinds for several megayears.

The same physics that is important in determining stellar wind bubble
dynamics may also act in planetary nebulae, as reviewed by Guerrero.
In a series of papers \citet{toala2014,toala2016,toala2018} showed
that thin-shell and Rayleigh-Taylor instabilities, along with
shadowing of ionizing radiation, would mix the contact discontinuity between hot
and cold gas in these systems as well, reducing X-ray emission
compared to pure thermal conduction.  This has been associated with an observed
increase in intermediate ions such as \ion{N}{5} at the discontinuity
\citep{fang2016}, and, as Richer pointed out, in broadening of UV lines
reaching 5--20~km~s$^{-1}$.

Wang noted that observational constraint of these mixing models can be
achieved using thermal plasma models of X-ray spectra that include
charge exchange \citep{zhang2014}. 
Because charge exchange is proportional to the ion flux into the
contact discontinuity, it can constrain the product of the flow speed
and the effective interface area produced by mixing.
\section{Coffee Break}

\begin{figure*}
  \includegraphics[width=\linewidth]{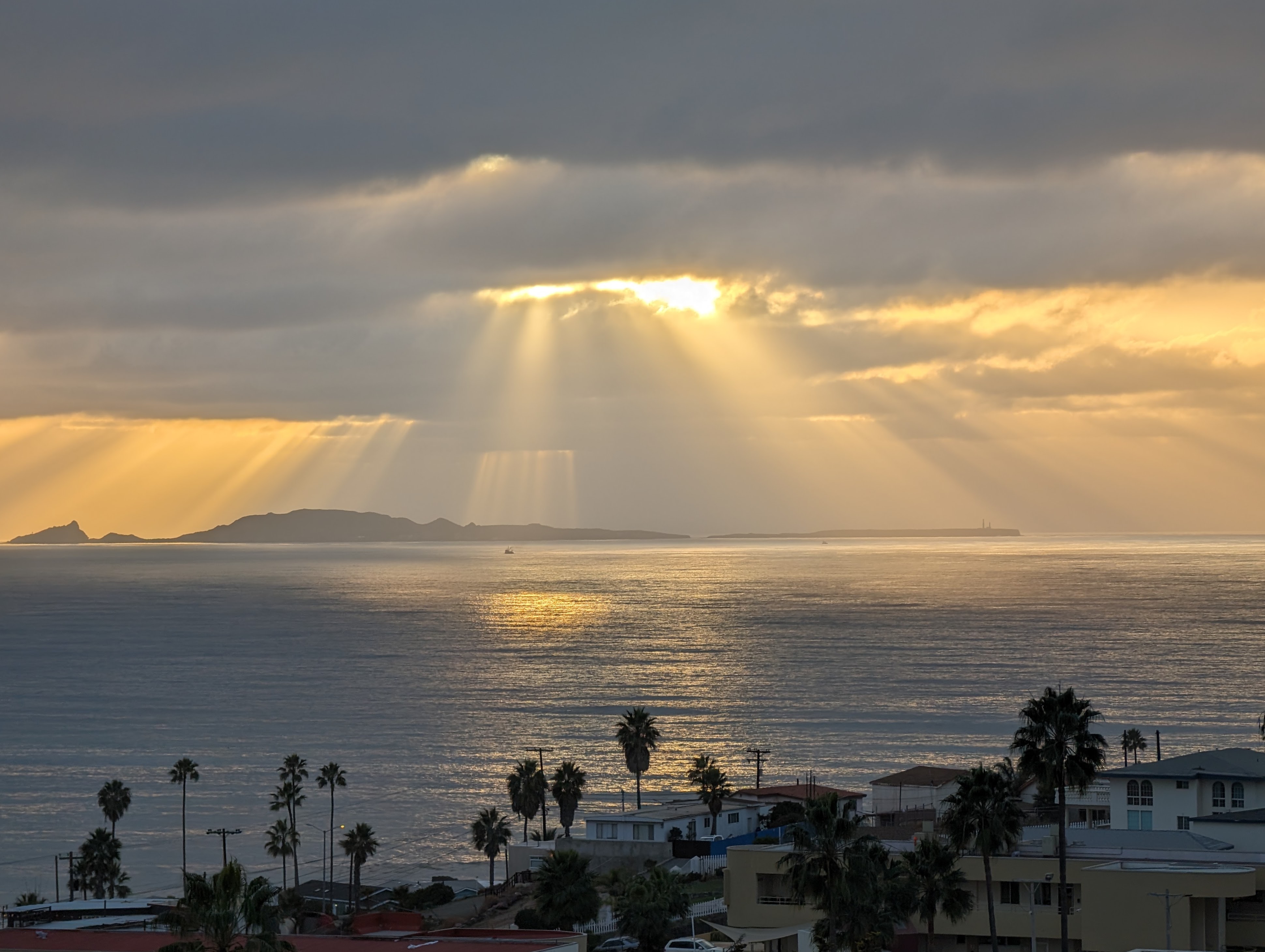}
  \caption{View from coffee break at IA-UNAM Ensenada.}
  \label{fig:coffee}
\end{figure*}

Coffee breaks at the IA-UNAM in Ensenada had a spectacular view, as
shown in Figure~\ref{fig:coffee}.

\section{Shiny Results}
In this section I pick out results reported at the conference that I
judged to be of particular interest, but do not
follow a single theme.

\subsection{Planet Formation}
Recent observations with the Atacama Large Millimeter/submillimeter
Array (ALMA) have dramatically sharpened our view of the early stages
of disk formation.  This period increasingly looks likely to be the
crucial period for planet formation.  Two major surveys have shown
stark differences in the appearance that disks show over the first
few megayears of their lives.  The eDISK survey \citep{eDISK} focused
on class 0 disks shows rather uniform disks, while the DSHARP survey
\citep{DSHARP} focused on class I and II disks famously shows a wide
variety of rings, spirals, gaps, and other structures.  The jury is
still out on whether the lack of structure at early times is an
optical depth effect---earlier, more massive disks might be too
optically thick to show midplane structure---or whether the
development of disk structure correlates with the growth of gas giant
planets. This is, of course, complicated by the argument on whether
disk structures are actually caused by planet formation or other
mechanisms, as only two planets, PDS 70b and PDS 70c, have actually
been observed in a disk with a gap \citep[e.g.][]{PDS70}.

An alternative proposal for producing structure in observed dust disks
is the impact of non-uniform accretion onto disks.  Segura-Cox
reviewed observational evidence for streamers of gas accreting
onto protoplanetary disks
\citep{alma2015,segura-cox2020,garufi2022, flores2023}. \citet{kuznetsova2022} has made
the argument that the impact of streamers on disks forms pressure
bumps that can trap gas, providing both a promising site for rapid
planet formation and an alternative explanation for the formation of
the observed ring structures.

Once disks form, dust settles to the midplance, where it begins to
coagulate into grains that can grow large enough to start decoupling
from the gas.  When stopping times grow towards orbital time scales
(Stokes number approaches unity), streaming instability sets in,
gathering particles into dense clumps that can become
self-gravitating.  Yang reviewed high-resolution models \citep{yang2014,schaefer2017} of the
streaming instability that show a remarkable resemblence to the size
distribution of the cold, classical Kuiper Belt observed by
\citet{kavelaars2021}.  This region of the Kuiper Belt is sufficiently
low density that the objects are expected to retain their primordial
size distribution, making it an excellent laboratory for study of
planetesimal formation.

\subsection{Star Formation}
Two models for star cluster formation were discussed.  Grebel reviewed
simulations showing that collisions of discrete spherical clouds with
different masses in the interstellar medium produce characteristic
U-shaped clouds with cavities morphologically similar to observed
H~{\sc ii} regions such as RCW 120, S44, or S36.  However, Arthur's
talk showed that the champagne flow morphology characteristic of
massive star formation in a region with a density gradient equally well
describes these regions.  V\'azquez-Semadeni argued that global
hierarchical collapse in a turbulent interstellar medium better
describes the star formation process.  Turbulent flows produce a
continuous density distribution poorly described by isolated, discrete
clouds, but easily leading to the density gradients needed to produced
U-shaped bubbles.

Another issue discussed was the structure of the magnetic fields that can
prevent or allow gravitational collapse and star formation.  On the scale of a filament
dozens of parsecs long, Stephens showed a magnetic polarization map
revealing that although the field lies predominantly perpendicular
to the filament, there are also multiple regions where it is
parallel.  This suggests that although the field is important to
shaping the flow, it does not always dominate.  Looney used
ALMA 870 $\mu$m polarization to demonstrate grain alignment in the
disk of HL Tau \citep{stephens2023}, presumably by the local field in
the disk.  Sharma compared polarization measurements towards the
outflow and envelope around HD 200775 taken with Planck and AIMPOL in
India, showing how the low-resolution, large-scale Planck results
average over the small-scale structure in the region.

\subsection{Nova and Supernova Remnants}
Orozco-Duarte reviewed the varied morphology of SN remnants,
and compared them to simulations of three typical scenarios: an
explosion within a bow shock produced by a star moving supersonically
with respect to the surrounding medium, a explosion within a star's
(spherically symmetric) birth cloud, and an explosion near the edge of a
filament an order of magnitude denser than the surrounding medium,
which allow reproduction of many observed SN morphologies.
\citet{orozco-duarte2023} showed that a superbubble in the filament
scenario will have off-center SN explosions that cleanly
explain the observed soft X-ray luminosity, supporting the hypothesis
originally proposed by \citet{chu1990}.

Santamaria and collaborators had a poster showing a morphological
catalog of nova remnants.  The frequent occurrence of fragmented
shells is striking.  \citet{toraskar2013} used simulations to
demonstrate that this is exactly the morphology expected from
repeating nova explosions separated by periods of hibernation.

A three-dimensional model of the Gemini-Monoceros X-ray enhancement
using eROSITA data \citep{knies2024} was reviewed by Sasaki.  Rather
than the usual conclusion that there are two overlapping remnants in
the region, they found a total of four objects overlapping in various ways.

The magnetic field in the region behind a SN blast wave was
also reviewed by Sasaki.  Evolving small scale structure oriented
perpendicularly to the blast wave was identified by
\citet{matsuda2020}.

Type Ia remnants were considered by several speakers.  Pan showed the
effects of Type Ia ejecta hitting companion stars \citep{pan2012a}, while Chuan-Jui Li
showed the effect of circumstellar medium around Type Ia SNe on
their remnants. \citet{cjli2021} showed evidence that the presence of circumstellar medium could be 
more common than expected, and derived an evolutionary
sequence for these remnants.

\subsection{Stars in all their variety}

\subsubsection{Moving Stars}
Arthur and Mackey both emphasized that stars with strong stellar winds moving through the ISM
produce distinctive bow shocks \citep{van_buren1992} and bow waves
\citep{henney2019a,henney2019b,henney2019c}. These can include wind bow shocks such as NGC 7635
\citep{green2019} or $\zeta$~Oph \citep{toala2016,green2019}, bow
waves such as the boundary of the heliosphere, dust waves, and
radiation bow shocks.  Wind bow shocks around stars with fast enough
winds can even be detected in the X-ray
\citep{toala2016,green2022}. Orozco-Duarte showed the consequences of
an SN explosion within a bow shock \citep{orozco-duarte2023}.

\subsubsection{Very Massive Stars}

The hunt for very massive stars with masses well above 100~M$_\odot$
has extended for decades.  The bright spot R136 at the center of
You-Hua's favorite H~{\sc ii} region, 30 Doradus, was already
hypothesized to be a single 1000~M$_\odot$ object in the early
1980s.  One of her early scientific successes was splitting that spot
into multiple components, demonstrating that it is a cluster and not a
single star \citep{chu1984}. However, Smith reviewed the evidence for
very massive stars with masses far in excess of 100~M$_\odot$
dominating the core of that cluster, most importantly the strong
He~{\sc ii} emission lines observed there \citep{crowther2016}. Other
super star clusters also show similar emission, arguing that very
massive stars are quite generally present in these exceptional
objects, even out to high redshift.  Wofford also reviewed this evidence in other objects, such
as NGC~3125-A1 \citep{martins2022}.

\subsubsection{Close Binary Evolution}

There was, of course, extensive discussion of the evolution of close
binaries, which can lead to anything from a planetary nebula to a
kilonova produced by a neutron star merger.  Ricker showed the results
of the evolution of a tight binary where mass transfer from the more
massive primary to the secondary prior to the SN explosion of
the primary results in the secondary evolving faster than it would
otherwise, allowing a common envelope to form in the envelope of the
secondary encompassing the neutron star remnant of the primary.  The
end result is a neutron star binary that can merge in a kilonova.
Estrada showed that mass transfer from a low mass star onto a compact
companion can strip enough mass away to leave a planetary-mass object,
which he dubbed a ``Chupiter''.

Garc\'{\i}a-Segura showed his increasingly detailed models of common
envelope evolution that now couple one-dimensional MESA models of
post-main sequence stellar evolution to three-dimensional Flash models
\citep[begun by][]{ricker2012} and aspherical two-dimensional ZEUS
models
\citep{garcia-segura2018,garcia-segura2020,garcia-segura2021,garcia-segura2022}.
The broad variety of planetary nebula morphologies can be captured by
this technique to a surprising extent.

Richer studied the velocity gradients and line broadening in planetary
nebula shells, showing that the ordered velocity gradient can not
explain the full line widths observed.  This suggests that turbulent
energy in the shell could be as much as 25\% of the thermal energy of
the plasma, something not accounted for in previous studies.
Weis used similar line observations of AG Carinae to demonstrate that
it is not elliptical, but instead bipolar, a morphology that is
obscured by its pole-on orientation towards the Earth.


Haberl showed observational evidence that the population of high-mass X-ray binaries
correlates well with the star formation rate 25--60~Myr prior in the
Small Milky Cloud \citep{antoniou2010}, but in the LMC
correlates with the rate 6--25~Myr prior, and with a formation
efficiency 17 times lower in the higher metallicity region
\citep{antoniou2016}.

\subsection{Star Clusters}

Stars clearly form in a non-uniform manner.  This was classically
thought of as occurring in two modes of star formation: clustered and
isolated.  Grebel emphasized in her review, however, the result of
\citet{bressert2010} that young stellar objects show a continuous Gaussian
distribution of surface densities with a peak in the region within 500
pc of the Sun of 22 pc$^{-2}$ and a dispersion in the log of the
surface density of 0.85.  Regions at the high end of this
distribution get identified as clusters, but the choice of a cutoff
between clustered and isolated star formation is arbitrary.

Similarly, the mass-radius relation of observed clusters appears to
show a continuous distribution from open clusters through globular
clusters, if one takes into account that young massive clusters can
have masses and radii intermediate between open clusters and old
globular clusters \citep{portegies-zwart2010}.
The evolution of clusters in the mass-radius plane can be seen in
action as recent observations \citep[e.g.][]{drew2019,meingast2021}
show that open clusters are often 
accompanied by enormous halos of unbound stars of the same age
occupying a region as much as an order of magnitude larger than the
tidal radius of the central cluster.


\subsection{Surveys}
Multiple observational surveys were reviewed.  Rodriguez described
the addition of {\em James Webb Space Telescope} data on 19 spiral galaxies
to the PHANGS survey of
nearby galaxies at high resolution.  Eight infrared bands were imaged
with the MIRI and NIRCAM instruments, providing access to stellar photospheric
emission with low obscuration, polycyclic aromatic hydrocarbons, dust
continuum, and silicate absorption.  Dale showed how the combination
of the LEGUS \citep{calzetti2015}, PHANGS
\citep{lee2022,leroy2021,emsellem2022} and GOALS
\citep{armus2009} surveys shows a relationship between stellar mass
and star formation rate spanning five orders of magnitude.

Maschmann
used the PHANGS-HST data to study the ages of clusters across the
mass-star formation rate plane.  Galaxies with high star formation rates
compared to the typical value (the so-called main sequence)
have plenty of middle-aged clusters, while galaxies with low rates
tend to be missing them.  (The figure showing this effect was a
contender for the most data presented in one figure, as color-color
diagrams for every galaxy were presented in a single mass-star
formation rate plot.)

Points described the Milky Clouds Emission Line Survey in its most
recent version using the Dark Energy Camera.  Williams used the Survey
to identify a large number of SN remnants across the LMC.  S\'anchez reviewed the Local Volume Mapper, which
uses integral field units the size of the full moon to take spectra at
30'' resolution sampling the full sky and densely covering the plane
of the Milky Way disk, as well as Orion and the Milky Clouds.

Haberl reviewed the eROSITA all-sky X-ray surveys, the first four of
which have been completed, and the fifth of which was truncated by
unfortunate geopolitical events, but not before covering the northern
half of the LMC.  Altogether, some LMC sources have as much as three
weeks of observation time.  Grishunin reviewd the APEX Legacy LMC
CO-line Survey, which gives 5 pc resolution across 85\% of the LMC,
resolving clouds with masses as low as 300~M$_\odot$.

\subsection{Numerical Techniques}

Several talks emphasized the need to pay close attention to numerical
issues to ensure that the physics is being captured.  Resolution of
physical length scales is a near universal issue.  Pittard showed a
quantitative criterion for how well the source region for a stellar
wind bubble must be resolved to ensure that a bubble forms at all, and
further with the correct radial momentum. Mackey showed that increasing
the resolution of a bow shock by a factor of four dramatically
increases the amount of mass entrained from the contact discontinuity
between shocked wind and swept-up ISM
by Kelvin-Helmholtz instabilities in the tail of the structure.
Mathew checked the ionization structure of an adiabatic shock across a
factor of 10 in linear resolution, finding that 1024 grid points does
a very good job.

\subsection{Observational Techniques}
The past and the future of the observation of bubbles and SN
remnants was discussed.  Toal\'a reminded us of the dramatic advance
in imaging capability represented by the transition from the {\em
  Einstein Observatory} to {\em XMM-Newton} using the example of
images of the stellar wind bubble S308.  He brought us up to the
present day with an infrared spectrum of extraordinary resolution of
WR124, taken with the {\em James Webb Space Telescope} MIRI integral
field unit.  Then, he compared the {\em XMM} spectrum of S308 to
simulated spectra expected from {\em XRISM}, {\em AXIS}, and finally,
and most extraordinarily, the exquisite spectrum out to 3 keV expected
from the {\em Athena} WFI.

Long emphasized that {\em XRISM} will be
able to take high-resolution spectra of Galactic SN remnants,
while {\em Athena} will extend that capability to nearby galaxies.
These spectra will allow measurement of rarer elements than possible
to date, constraining explosion mechanisms.  {\em IXPE} will measure
the polarization of Galactic SN remnants, constraining their
field structures and thus their particle acceleration properties.
Meerkat and upcoming radio telescopes can now image remnants at radio
wavelengths with the angular resolution we are accustomed to from
optical observatories.  This will allow discovery and characterization
of remnants in other galaxies.

\section{Recognition}
The scientific organizing committee of this conference felt that the
best recognition we could make of You-Hua Chu's scientific career was
to include her in Wikipedia.  The stringent standards currently
applied for notability of entries allowed into this encyclopedia
indeed show the importance of her career.  See
\texttt{en.wikipedia.org/wiki/You-Hua\_Chu} for the current version of
this page.  A page has since also been
added in simplified Chinese
(\texttt{zh.wikipedia.org/wiki/\%E6\%9C\%B1\%E6\%9C\%89\%E8\%8A\%B1}).



\begin{thebibliography}
\expandafter\ifx\csname natexlab\endcsname\relax\def\natexlab#1{#1}\fi
\expandafter\ifx\csname href\endcsname\relax
  \def\href#1#2{}\fi
\expandafter\ifx\csname urllinklabel\endcsname\relax
  \def\urllinklabel{[LINK]}\fi
\expandafter\ifx\csname adsurllinklabel\endcsname\relax
  \def\adsurllinklabel{[ADS]}\fi

\bibitem[{{ALMA Partnership} {et~al.}(2015){ALMA Partnership}, {Brogan},
  {P{\'e}rez}, {Hunter}, {Dent}, {Hales}, {Hills}, {Corder}, {Fomalont},
  {Vlahakis}, {Asaki}, {Barkats}, {Hirota}, {Hodge}, {Impellizzeri}, {Kneissl},
  {Liuzzo}, {Lucas}, {Marcelino}, {Matsushita}, {Nakanishi}, {Phillips},
  {Richards}, {Toledo}, {Aladro}, {Broguiere}, {Cortes}, {Cortes}, {Espada},
  {Galarza}, {Garcia-Appadoo}, {Guzman-Ramirez}, {Humphreys}, {Jung}, {Kameno},
  {Laing}, {Leon}, {Marconi}, {Mignano}, {Nikolic}, {Nyman}, {Radiszcz},
  {Remijan}, {Rod{\'o}n}, {Sawada}, {Takahashi}, {Tilanus}, {Vila Vilaro},
  {Watson}, {Wiklind}, {Akiyama}, {Chapillon}, {de Gregorio-Monsalvo}, {Di
  Francesco}, {Gueth}, {Kawamura}, {Lee}, {Nguyen Luong}, {Mangum}, {Pietu},
  {Sanhueza}, {Saigo}, {Takakuwa}, {Ubach}, {van Kempen}, {Wootten},
  {Castro-Carrizo}, {Francke}, {Gallardo}, {Garcia}, {Gonzalez}, {Hill},
  {Kaminski}, {Kurono}, {Liu}, {Lopez}, {Morales}, {Plarre}, {Schieven},
  {Testi}, {Videla}, {Villard}, {Andreani}, {Hibbard}, \&
  {Tatematsu}}]{alma2015}
{ALMA Partnership}, {Brogan}, C.~L., {P{\'e}rez}, L.~M., {Hunter}, T.~R.,
  {Dent}, W.~R.~F., {Hales}, A.~S., {Hills}, R.~E., {Corder}, S., {Fomalont},
  E.~B., {Vlahakis}, C., {Asaki}, Y., {Barkats}, D., {Hirota}, A., {Hodge},
  J.~A., {Impellizzeri}, C.~M.~V., {Kneissl}, R., {Liuzzo}, E., {Lucas}, R.,
  {Marcelino}, N., {Matsushita}, S., {Nakanishi}, K., {Phillips}, N.,
  {Richards}, A.~M.~S., {Toledo}, I., {Aladro}, R., {Broguiere}, D., {Cortes},
  J.~R., {Cortes}, P.~C., {Espada}, D., {Galarza}, F., {Garcia-Appadoo}, D.,
  {Guzman-Ramirez}, L., {Humphreys}, E.~M., {Jung}, T., {Kameno}, S., {Laing},
  R.~A., {Leon}, S., {Marconi}, G., {Mignano}, A., {Nikolic}, B., {Nyman},
  L.~A., {Radiszcz}, M., {Remijan}, A., {Rod{\'o}n}, J.~A., {Sawada}, T.,
  {Takahashi}, S., {Tilanus}, R.~P.~J., {Vila Vilaro}, B., {Watson}, L.~C.,
  {Wiklind}, T., {Akiyama}, E., {Chapillon}, E., {de Gregorio-Monsalvo}, I.,
  {Di Francesco}, J., {Gueth}, F., {Kawamura}, A., {Lee}, C.~F., {Nguyen
  Luong}, Q., {Mangum}, J., {Pietu}, V., {Sanhueza}, P., {Saigo}, K.,
  {Takakuwa}, S., {Ubach}, C., {van Kempen}, T., {Wootten}, A.,
  {Castro-Carrizo}, A., {Francke}, H., {Gallardo}, J., {Garcia}, J.,
  {Gonzalez}, S., {Hill}, T., {Kaminski}, T., {Kurono}, Y., {Liu}, H.~Y.,
  {Lopez}, C., {Morales}, F., {Plarre}, K., {Schieven}, G., {Testi}, L.,
  {Videla}, L., {Villard}, E., {Andreani}, P., {Hibbard}, J.~E., \&
  {Tatematsu}, K. 2015, \apjl, 808, L3


\bibitem[{Andrews {et~al.}(2018)Andrews, Huang, P\'erez, Isella, Dullemond,
  Kurtovic, Guzm\'an, Carpenter, Wilner, Zhang, Zhu, Birnstiel, Bai, Benisty,
  Hughes, \"Oberg, \& Ricci}]{DSHARP}
Andrews, S.~M., Huang, J., P\'erez, L.~M., Isella, A., Dullemond, C.~P.,
  Kurtovic, N.~T., Guzm\'an, V.~V., Carpenter, J.~M., Wilner, D.~J., Zhang, S.,
  Zhu, Z., Birnstiel, T., Bai, X.-N., Benisty, M., Hughes, A.~M., \"Oberg,
  K.~I., \& Ricci, L. 2018, The Astrophysical Journal Letters, 869, L41
 \href{https://dx.doi.org/10.3847/2041-8213/aaf741}{\urllinklabel}

\bibitem[{{Antoniou} \& {Zezas}(2016)}]{antoniou2016}
{Antoniou}, V. \& {Zezas}, A. 2016, \mnras, 459, 528


\bibitem[{{Antoniou} {et~al.}(2010){Antoniou}, {Zezas}, {Hatzidimitriou}, \&
  {Kalogera}}]{antoniou2010}
{Antoniou}, V., {Zezas}, A., {Hatzidimitriou}, D., \& {Kalogera}, V. 2010,
  \apjl, 716, L140


\bibitem[{{Armus} {et~al.}(2009){Armus}, {Mazzarella}, {Evans}, {Surace},
  {Sanders}, {Iwasawa}, {Frayer}, {Howell}, {Chan}, {Petric}, {Vavilkin},
  {Kim}, {Haan}, {Inami}, {Murphy}, {Appleton}, {Barnes}, {Bothun}, {Bridge},
  {Charmandaris}, {Jensen}, {Kewley}, {Lord}, {Madore}, {Marshall},
  {Melbourne}, {Rich}, {Satyapal}, {Schulz}, {Spoon}, {Sturm}, {U}, {Veilleux},
  \& {Xu}}]{armus2009}
{Armus}, L., {Mazzarella}, J.~M., {Evans}, A.~S., {Surace}, J.~A., {Sanders},
  D.~B., {Iwasawa}, K., {Frayer}, D.~T., {Howell}, J.~H., {Chan}, B., {Petric},
  A., {Vavilkin}, T., {Kim}, D.~C., {Haan}, S., {Inami}, H., {Murphy}, E.~J.,
  {Appleton}, P.~N., {Barnes}, J.~E., {Bothun}, G., {Bridge}, C.~R.,
  {Charmandaris}, V., {Jensen}, J.~B., {Kewley}, L.~J., {Lord}, S., {Madore},
  B.~F., {Marshall}, J.~A., {Melbourne}, J.~E., {Rich}, J., {Satyapal}, S.,
  {Schulz}, B., {Spoon}, H.~W.~W., {Sturm}, E., {U}, V., {Veilleux}, S., \&
  {Xu}, K. 2009, \pasp, 121, 559


\bibitem[{{Avedisova}(1972)}]{avedisova1972}
{Avedisova}, V.~S. 1972, \sovast, 15, 708


\bibitem[{{Benisty} {et~al.}(2021){Benisty}, {Bae}, {Facchini}, {Keppler},
  {Teague}, {Isella}, {Kurtovic}, {P{\'e}rez}, {Sierra}, {Andrews},
  {Carpenter}, {Czekala}, {Dominik}, {Henning}, {Menard}, {Pinilla}, \&
  {Zurlo}}]{PDS70}
{Benisty}, M., {Bae}, J., {Facchini}, S., {Keppler}, M., {Teague}, R.,
  {Isella}, A., {Kurtovic}, N.~T., {P{\'e}rez}, L.~M., {Sierra}, A., {Andrews},
  S.~M., {Carpenter}, J., {Czekala}, I., {Dominik}, C., {Henning}, T.,
  {Menard}, F., {Pinilla}, P., \& {Zurlo}, A. 2021, \apjl, 916, L2


\bibitem[{{Bressert} {et~al.}(2010){Bressert}, {Bastian}, {Gutermuth},
  {Megeath}, {Allen}, {Evans}, {Rebull}, {Hatchell}, {Johnstone}, {Bourke},
  {Cieza}, {Harvey}, {Merin}, {Ray}, \& {Tothill}}]{bressert2010}
{Bressert}, E., {Bastian}, N., {Gutermuth}, R., {Megeath}, S.~T., {Allen}, L.,
  {Evans}, II, N.~J., {Rebull}, L.~M., {Hatchell}, J., {Johnstone}, D.,
  {Bourke}, T.~L., {Cieza}, L.~A., {Harvey}, P.~M., {Merin}, B., {Ray}, T.~P.,
  \& {Tothill}, N.~F.~H. 2010, \mnras, 409, L54


\bibitem[{{Brown} {et~al.}(1995){Brown}, {Hartmann}, \& {Burton}}]{brown1995}
{Brown}, A.~G.~A., {Hartmann}, D., \& {Burton}, W.~B. 1995, \aap, 300, 903


\bibitem[{{Calzetti} {et~al.}(2015){Calzetti}, {Lee}, {Sabbi}, {Adamo},
  {Smith}, {Andrews}, {Ubeda}, {Bright}, {Thilker}, {Aloisi}, {Brown},
  {Chandar}, {Christian}, {Cignoni}, {Clayton}, {da Silva}, {de Mink}, {Dobbs},
  {Elmegreen}, {Elmegreen}, {Evans}, {Fumagalli}, {Gallagher}, {Gouliermis},
  {Grebel}, {Herrero}, {Hunter}, {Johnson}, {Kennicutt}, {Kim}, {Krumholz},
  {Lennon}, {Levay}, {Martin}, {Nair}, {Nota}, {{\"O}stlin}, {Pellerin},
  {Prieto}, {Regan}, {Ryon}, {Schaerer}, {Schiminovich}, {Tosi}, {Van Dyk},
  {Walterbos}, {Whitmore}, \& {Wofford}}]{calzetti2015}
{Calzetti}, D., {Lee}, J.~C., {Sabbi}, E., {Adamo}, A., {Smith}, L.~J.,
  {Andrews}, J.~E., {Ubeda}, L., {Bright}, S.~N., {Thilker}, D., {Aloisi}, A.,
  {Brown}, T.~M., {Chandar}, R., {Christian}, C., {Cignoni}, M., {Clayton},
  G.~C., {da Silva}, R., {de Mink}, S.~E., {Dobbs}, C., {Elmegreen}, B.~G.,
  {Elmegreen}, D.~M., {Evans}, A.~S., {Fumagalli}, M., {Gallagher}, J.~S., I.,
  {Gouliermis}, D.~A., {Grebel}, E.~K., {Herrero}, A., {Hunter}, D.~A.,
  {Johnson}, K.~E., {Kennicutt}, R.~C., {Kim}, H., {Krumholz}, M.~R., {Lennon},
  D., {Levay}, K., {Martin}, C., {Nair}, P., {Nota}, A., {{\"O}stlin}, G.,
  {Pellerin}, A., {Prieto}, J., {Regan}, M.~W., {Ryon}, J.~E., {Schaerer}, D.,
  {Schiminovich}, D., {Tosi}, M., {Van Dyk}, S.~D., {Walterbos}, R.,
  {Whitmore}, B.~C., \& {Wofford}, A. 2015, \aj, 149, 51


\bibitem[{{Castor} {et~al.}(1975){Castor}, {McCray}, \& {Weaver}}]{castor1975}
{Castor}, J., {McCray}, R., \& {Weaver}, R. 1975, \apjl, 200, L107


\bibitem[{{Chu} {et~al.}(1984){Chu}, {Cassinelli}, \& {Wolfire}}]{chu1984}
{Chu}, Y.~H., {Cassinelli}, J.~P., \& {Wolfire}, M.~G. 1984, \apj, 283, 560


\bibitem[{{Chu} \& {Mac Low}(1990)}]{chu1990}
{Chu}, Y.-H. \& {Mac Low}, M.-M. 1990, \apj, 365, 510


\bibitem[{{Cooper} {et~al.}(2004){Cooper}, {Guerrero}, {Chu}, {Chen}, \&
  {Dunne}}]{cooper2004}
{Cooper}, R.~L., {Guerrero}, M.~A., {Chu}, Y.-H., {Chen}, C. H.~R., \& {Dunne},
  B.~C. 2004, \apj, 605, 751


\bibitem[{{Crowther} {et~al.}(2016){Crowther}, {Caballero-Nieves}, {Bostroem},
  {Ma{\'\i}z Apell{\'a}niz}, {Schneider}, {Walborn}, {Angus}, {Brott},
  {Bonanos}, {de Koter}, {de Mink}, {Evans}, {Gr{\"a}fener}, {Herrero},
  {Howarth}, {Langer}, {Lennon}, {Puls}, {Sana}, \& {Vink}}]{crowther2016}
{Crowther}, P.~A., {Caballero-Nieves}, S.~M., {Bostroem}, K.~A., {Ma{\'\i}z
  Apell{\'a}niz}, J., {Schneider}, F.~R.~N., {Walborn}, N.~R., {Angus}, C.~R.,
  {Brott}, I., {Bonanos}, A., {de Koter}, A., {de Mink}, S.~E., {Evans}, C.~J.,
  {Gr{\"a}fener}, G., {Herrero}, A., {Howarth}, I.~D., {Langer}, N., {Lennon},
  D.~J., {Puls}, J., {Sana}, H., \& {Vink}, J.~S. 2016, \mnras, 458, 624


\bibitem[{{Dalla Vecchia} \& {Schaye}(2008)}]{dalla_vecchia2008}
{Dalla Vecchia}, C. \& {Schaye}, J. 2008, \mnras, 387, 1431


\bibitem[{{Danehkar} {et~al.}(2021){Danehkar}, {Oey}, \& {Gray}}]{danehkar2021}
{Danehkar}, A., {Oey}, M.~S., \& {Gray}, W.~J. 2021, \apj, 921, 91


\bibitem[{{Drew} {et~al.}(2019){Drew}, {Mongui{\'o}}, \& {Wright}}]{drew2019}
{Drew}, J.~E., {Mongui{\'o}}, M., \& {Wright}, N.~J. 2019, \mnras, 486, 1034


\bibitem[{{Dyson}(1975)}]{dyson1975}
{Dyson}, J.~E. 1975, \apss, 35, 299


\bibitem[{{Emsellem} {et~al.}(2022){Emsellem}, {Schinnerer}, {Santoro},
  {Belfiore}, {Pessa}, {McElroy}, {Blanc}, {Congiu}, {Groves}, {Ho}, {Kreckel},
  {Razza}, {Sanchez-Blazquez}, {Egorov}, {Faesi}, {Klessen}, {Leroy}, {Meidt},
  {Querejeta}, {Rosolowsky}, {Scheuermann}, {Anand}, {Barnes},
  {Be{\v{s}}li{\'c}}, {Bigiel}, {Boquien}, {Cao}, {Chevance}, {Dale},
  {Eibensteiner}, {Glover}, {Grasha}, {Henshaw}, {Hughes}, {Koch}, {Kruijssen},
  {Lee}, {Liu}, {Pan}, {Pety}, {Saito}, {Sandstrom}, {Schruba}, {Sun},
  {Thilker}, {Usero}, {Watkins}, \& {Williams}}]{emsellem2022}
{Emsellem}, E., {Schinnerer}, E., {Santoro}, F., {Belfiore}, F., {Pessa}, I.,
  {McElroy}, R., {Blanc}, G.~A., {Congiu}, E., {Groves}, B., {Ho}, I.~T.,
  {Kreckel}, K., {Razza}, A., {Sanchez-Blazquez}, P., {Egorov}, O., {Faesi},
  C., {Klessen}, R.~S., {Leroy}, A.~K., {Meidt}, S., {Querejeta}, M.,
  {Rosolowsky}, E., {Scheuermann}, F., {Anand}, G.~S., {Barnes}, A.~T.,
  {Be{\v{s}}li{\'c}}, I., {Bigiel}, F., {Boquien}, M., {Cao}, Y., {Chevance},
  M., {Dale}, D.~A., {Eibensteiner}, C., {Glover}, S. C.~O., {Grasha}, K.,
  {Henshaw}, J.~D., {Hughes}, A., {Koch}, E.~W., {Kruijssen}, J.~M.~D., {Lee},
  J., {Liu}, D., {Pan}, H.-A., {Pety}, J., {Saito}, T., {Sandstrom}, K.~M.,
  {Schruba}, A., {Sun}, J., {Thilker}, D.~A., {Usero}, A., {Watkins}, E.~J., \&
  {Williams}, T.~G. 2022, \aap, 659, A191


\bibitem[{{Fang} {et~al.}(2016){Fang}, {Guerrero}, {Toal\'a}, {Chu}, \&
  {Gruendl}}]{fang2016}
{Fang}, X., {Guerrero}, M.~A., {Toal\'a}, J.~A., {Chu}, Y.-H., \& {Gruendl},
  R.~A. 2016, \apjl, 822, L19


\bibitem[{{Flores} {et~al.}(2023){Flores}, {Ohashi}, {Tobin}, {J{\o}rgensen},
  {Takakuwa}, {Li}, {Lin}, {van't Hoff}, {Plunkett}, {Yamato}, {Sai (Insa
  Choi)}, {Koch}, {Yen}, {Aikawa}, {Aso}, {de Gregorio-Monsalvo}, {Kido},
  {Kwon}, {Lee}, {Lee}, {Looney}, {Santamar{\'\i}a-Miranda}, {Sharma},
  {Thieme}, {Williams}, {Han}, {Narayanan}, \& {Lai}}]{flores2023}
{Flores}, C., {Ohashi}, N., {Tobin}, J.~J., {J{\o}rgensen}, J.~K., {Takakuwa},
  S., {Li}, Z.-Y., {Lin}, Z.-Y.~D., {van't Hoff}, M. L.~R., {Plunkett}, A.~L.,
  {Yamato}, Y., {Sai (Insa Choi)}, J., {Koch}, P.~M., {Yen}, H.-W., {Aikawa},
  Y., {Aso}, Y., {de Gregorio-Monsalvo}, I., {Kido}, M., {Kwon}, W., {Lee},
  J.-E., {Lee}, C.~W., {Looney}, L.~W., {Santamar{\'\i}a-Miranda}, A.,
  {Sharma}, R., {Thieme}, T.~J., {Williams}, J.~P., {Han}, I., {Narayanan}, S.,
  \& {Lai}, S.-P. 2023, \apj, 958, 98


\bibitem[{{Freyer} {et~al.}(2003){Freyer}, {Hensler}, \& {Yorke}}]{freyer2003}
{Freyer}, T., {Hensler}, G., \& {Yorke}, H.~W. 2003, \apj, 594, 888


\bibitem[{{Garc{\'\i}a-Segura} {et~al.}(2018){Garc{\'\i}a-Segura}, {Ricker}, \&
  {Taam}}]{garcia-segura2018}
{Garc{\'\i}a-Segura}, G., {Ricker}, P.~M., \& {Taam}, R.~E. 2018, \apj, 860, 19


\bibitem[{{Garc{\'\i}a-Segura} {et~al.}(2020){Garc{\'\i}a-Segura}, {Taam}, \&
  {Ricker}}]{garcia-segura2020}
{Garc{\'\i}a-Segura}, G., {Taam}, R.~E., \& {Ricker}, P.~M. 2020, \apj, 893,
  150


\bibitem[{{Garc{\'\i}a-Segura} {et~al.}(2021){Garc{\'\i}a-Segura}, {Taam}, \&
  {Ricker}}]{garcia-segura2021}
---. 2021, \apj, 914, 111


\bibitem[{{Garc{\'\i}a-Segura} {et~al.}(2022){Garc{\'\i}a-Segura}, {Taam}, \&
  {Ricker}}]{garcia-segura2022}
---. 2022, \mnras, 517, 3822


\bibitem[{{Garufi} {et~al.}(2022){Garufi}, {Podio}, {Codella}, {Segura-Cox},
  {Vander Donckt}, {Mercimek}, {Bacciotti}, {Fedele}, {Kasper}, {Pineda},
  {Humphreys}, \& {Testi}}]{garufi2022}
{Garufi}, A., {Podio}, L., {Codella}, C., {Segura-Cox}, D., {Vander Donckt},
  M., {Mercimek}, S., {Bacciotti}, F., {Fedele}, D., {Kasper}, M., {Pineda},
  J.~E., {Humphreys}, E., \& {Testi}, L. 2022, \aap, 658, A104


\bibitem[{{Geen} {et~al.}(2015){Geen}, {Rosdahl}, {Blaizot}, {Devriendt}, \&
  {Slyz}}]{geen2015}
{Geen}, S., {Rosdahl}, J., {Blaizot}, J., {Devriendt}, J., \& {Slyz}, A. 2015,
  \mnras, 448, 3248


\bibitem[{{Green} {et~al.}(2019){Green}, {Mackey}, {Haworth}, {Gvaramadze}, \&
  {Duffy}}]{green2019}
{Green}, S., {Mackey}, J., {Haworth}, T.~J., {Gvaramadze}, V.~V., \& {Duffy},
  P. 2019, \aap, 625, A4


\bibitem[{{Green} {et~al.}(2022){Green}, {Mackey}, {Kavanagh}, {Haworth},
  {Moutzouri}, \& {Gvaramadze}}]{green2022}
{Green}, S., {Mackey}, J., {Kavanagh}, P., {Haworth}, T.~J., {Moutzouri}, M.,
  \& {Gvaramadze}, V.~V. 2022, \aap, 665, A35


\bibitem[{{Haid} {et~al.}(2018){Haid}, {Walch}, {Seifried}, {W{\"u}nsch},
  {Dinnbier}, \& {Naab}}]{haid2018}
{Haid}, S., {Walch}, S., {Seifried}, D., {W{\"u}nsch}, R., {Dinnbier}, F., \&
  {Naab}, T. 2018, \mnras, 478, 4799


\bibitem[{{Harper-Clark} \& {Murray}(2009)}]{harper-clark2009}
{Harper-Clark}, E. \& {Murray}, N. 2009, \apj, 693, 1696


\bibitem[{{Heger} {et~al.}(2003){Heger}, {Fryer}, {Woosley}, {Langer}, \&
  {Hartmann}}]{heger2003}
{Heger}, A., {Fryer}, C.~L., {Woosley}, S.~E., {Langer}, N., \& {Hartmann},
  D.~H. 2003, \apj, 591, 288


\bibitem[{{Henney} \& {Arthur}(2019{\natexlab{a}})}]{henney2019a}
{Henney}, W.~J. \& {Arthur}, S.~J. 2019{\natexlab{a}}, \mnras, 486, 3423


\bibitem[{{Henney} \& {Arthur}(2019{\natexlab{b}})}]{henney2019b}
---. 2019{\natexlab{b}}, \mnras, 486, 4423


\bibitem[{{Henney} \& {Arthur}(2019{\natexlab{c}})}]{henney2019c}
---. 2019{\natexlab{c}}, \mnras, 489, 2142


\bibitem[{{Hopkins} {et~al.}(2018){Hopkins}, {Wetzel}, {Kere{\v{s}}},
  {Faucher-Gigu{\`e}re}, {Quataert}, {Boylan-Kolchin}, {Murray}, {Hayward},
  {Garrison-Kimmel}, {Hummels}, {Feldmann}, {Torrey}, {Ma},
  {Angl{\'e}s-Alc{\'a}zar}, {Su}, {Orr}, {Schmitz}, {Escala}, {Sanderson},
  {Grudi{\'c}}, {Hafen}, {Kim}, {Fitts}, {Bullock}, {Wheeler}, {Chan},
  {Elbert}, \& {Narayanan}}]{hopkins2018}
{Hopkins}, P.~F., {Wetzel}, A., {Kere{\v{s}}}, D., {Faucher-Gigu{\`e}re},
  C.-A., {Quataert}, E., {Boylan-Kolchin}, M., {Murray}, N., {Hayward}, C.~C.,
  {Garrison-Kimmel}, S., {Hummels}, C., {Feldmann}, R., {Torrey}, P., {Ma}, X.,
  {Angl{\'e}s-Alc{\'a}zar}, D., {Su}, K.-Y., {Orr}, M., {Schmitz}, D.,
  {Escala}, I., {Sanderson}, R., {Grudi{\'c}}, M.~Y., {Hafen}, Z., {Kim},
  J.-H., {Fitts}, A., {Bullock}, J.~S., {Wheeler}, C., {Chan}, T.~K., {Elbert},
  O.~D., \& {Narayanan}, D. 2018, \mnras, 480, 800


\bibitem[{{Jaskot} {et~al.}(2019){Jaskot}, {Dowd}, {Oey}, {Scarlata}, \&
  {McKinney}}]{jaskot2019}
{Jaskot}, A.~E., {Dowd}, T., {Oey}, M.~S., {Scarlata}, C., \& {McKinney}, J.
  2019, \apj, 885, 96


\bibitem[{{Jecmen} \& {Oey}(2023)}]{jecmen2023}
{Jecmen}, M.~C. \& {Oey}, M.~S. 2023, \apj, 958, 149


\bibitem[{{Kavelaars} {et~al.}(2021){Kavelaars}, {Petit}, {Gladman},
  {Bannister}, {Alexandersen}, {Chen}, {Gwyn}, \& {Volk}}]{kavelaars2021}
{Kavelaars}, J.~J., {Petit}, J.-M., {Gladman}, B., {Bannister}, M.~T.,
  {Alexandersen}, M., {Chen}, Y.-T., {Gwyn}, S. D.~J., \& {Volk}, K. 2021,
  \apjl, 920, L28


\bibitem[{{Kim} \& {Ostriker}(2015)}]{kim2015}
{Kim}, C.-G. \& {Ostriker}, E.~C. 2015, \apj, 802, 99


\bibitem[{{Knies} {et~al.}(2024){Knies}, {Sasaki}, {Becker}, {Liu}, {Ponti}, \&
  {Plucinsky}}]{knies2024}
{Knies}, J.~R., {Sasaki}, M., {Becker}, W., {Liu}, T., {Ponti}, G., \&
  {Plucinsky}, P.~P. 2024, \aap, 688, A90


\bibitem[{{Komarova} {et~al.}(2021){Komarova}, {Oey}, {Krumholz}, {Silich},
  {Kumari}, \& {James}}]{komarova2021}
{Komarova}, L., {Oey}, M.~S., {Krumholz}, M.~R., {Silich}, S., {Kumari}, N., \&
  {James}, B.~L. 2021, \apjl, 920, L46


\bibitem[{{Krause} {et~al.}(2012){Krause}, {Charbonnel}, {Decressin}, {Meynet},
  {Prantzos}, \& {Diehl}}]{krause2012}
{Krause}, M., {Charbonnel}, C., {Decressin}, T., {Meynet}, G., {Prantzos}, N.,
  \& {Diehl}, R. 2012, \aap, 546, L5


\bibitem[{{Krumholz} {et~al.}(2009){Krumholz}, {McKee}, \&
  {Tumlinson}}]{krumholz2009}
{Krumholz}, M.~R., {McKee}, C.~F., \& {Tumlinson}, J. 2009, \apj, 699, 850


\bibitem[{{Kuznetsova} {et~al.}(2022){Kuznetsova}, {Bae}, {Hartmann}, \&
  {Low}}]{kuznetsova2022}
{Kuznetsova}, A., {Bae}, J., {Hartmann}, L., \& {Low}, M.-M.~M. 2022, \apj,
  928, 92


\bibitem[{{Lancaster} {et~al.}(2021){Lancaster}, {Ostriker}, {Kim}, \&
  {Kim}}]{lancaster2021}
{Lancaster}, L., {Ostriker}, E.~C., {Kim}, J.-G., \& {Kim}, C.-G. 2021, \apj,
  914, 90


\bibitem[{{Lee} {et~al.}(2022){Lee}, {Whitmore}, {Thilker}, {Deger}, {Larson},
  {Ubeda}, {Anand}, {Boquien}, {Chandar}, {Dale}, {Emsellem}, {Leroy},
  {Rosolowsky}, {Schinnerer}, {Schmidt}, {Lilly}, {Turner}, {Van Dyk}, {White},
  {Barnes}, {Belfiore}, {Bigiel}, {Blanc}, {Cao}, {Chevance}, {Congiu},
  {Egorov}, {Glover}, {Grasha}, {Groves}, {Henshaw}, {Hughes}, {Klessen},
  {Koch}, {Kreckel}, {Kruijssen}, {Liu}, {Lopez}, {Mayker}, {Meidt}, {Murphy},
  {Pan}, {Pety}, {Querejeta}, {Razza}, {Saito}, {S{\'a}nchez-Bl{\'a}zquez},
  {Santoro}, {Sardone}, {Scheuermann}, {Schruba}, {Sun}, {Usero}, {Watkins}, \&
  {Williams}}]{lee2022}
{Lee}, J.~C., {Whitmore}, B.~C., {Thilker}, D.~A., {Deger}, S., {Larson},
  K.~L., {Ubeda}, L., {Anand}, G.~S., {Boquien}, M., {Chandar}, R., {Dale},
  D.~A., {Emsellem}, E., {Leroy}, A.~K., {Rosolowsky}, E., {Schinnerer}, E.,
  {Schmidt}, J., {Lilly}, J., {Turner}, J., {Van Dyk}, S., {White}, R.~L.,
  {Barnes}, A.~T., {Belfiore}, F., {Bigiel}, F., {Blanc}, G.~A., {Cao}, Y.,
  {Chevance}, M., {Congiu}, E., {Egorov}, O.~V., {Glover}, S. C.~O., {Grasha},
  K., {Groves}, B., {Henshaw}, J.~D., {Hughes}, A., {Klessen}, R.~S., {Koch},
  E., {Kreckel}, K., {Kruijssen}, J.~M.~D., {Liu}, D., {Lopez}, L.~A.,
  {Mayker}, N., {Meidt}, S.~E., {Murphy}, E.~J., {Pan}, H.-A., {Pety}, J.,
  {Querejeta}, M., {Razza}, A., {Saito}, T., {S{\'a}nchez-Bl{\'a}zquez}, P.,
  {Santoro}, F., {Sardone}, A., {Scheuermann}, F., {Schruba}, A., {Sun}, J.,
  {Usero}, A., {Watkins}, E., \& {Williams}, T.~G. 2022, \apjs, 258, 10


\bibitem[{{Leroy} {et~al.}(2021){Leroy}, {Schinnerer}, {Hughes}, {Rosolowsky},
  {Pety}, {Schruba}, {Usero}, {Blanc}, {Chevance}, {Emsellem}, {Faesi},
  {Herrera}, {Liu}, {Meidt}, {Querejeta}, {Saito}, {Sandstrom}, {Sun},
  {Williams}, {Anand}, {Barnes}, {Behrens}, {Belfiore}, {Benincasa},
  {Be{\v{s}}li{\'c}}, {Bigiel}, {Bolatto}, {den Brok}, {Cao}, {Chandar},
  {Chastenet}, {Chiang}, {Congiu}, {Dale}, {Deger}, {Eibensteiner}, {Egorov},
  {Garc{\'\i}a-Rodr{\'\i}guez}, {Glover}, {Grasha}, {Henshaw}, {Ho}, {Kepley},
  {Kim}, {Klessen}, {Kreckel}, {Koch}, {Kruijssen}, {Larson}, {Lee}, {Lopez},
  {Machado}, {Mayker}, {McElroy}, {Murphy}, {Ostriker}, {Pan}, {Pessa},
  {Puschnig}, {Razza}, {S{\'a}nchez-Bl{\'a}zquez}, {Santoro}, {Sardone},
  {Scheuermann}, {Sliwa}, {Sormani}, {Stuber}, {Thilker}, {Turner}, {Utomo},
  {Watkins}, \& {Whitmore}}]{leroy2021}
{Leroy}, A.~K., {Schinnerer}, E., {Hughes}, A., {Rosolowsky}, E., {Pety}, J.,
  {Schruba}, A., {Usero}, A., {Blanc}, G.~A., {Chevance}, M., {Emsellem}, E.,
  {Faesi}, C.~M., {Herrera}, C.~N., {Liu}, D., {Meidt}, S.~E., {Querejeta}, M.,
  {Saito}, T., {Sandstrom}, K.~M., {Sun}, J., {Williams}, T.~G., {Anand},
  G.~S., {Barnes}, A.~T., {Behrens}, E.~A., {Belfiore}, F., {Benincasa}, S.~M.,
  {Be{\v{s}}li{\'c}}, I., {Bigiel}, F., {Bolatto}, A.~D., {den Brok}, J.~S.,
  {Cao}, Y., {Chandar}, R., {Chastenet}, J., {Chiang}, I.-D., {Congiu}, E.,
  {Dale}, D.~A., {Deger}, S., {Eibensteiner}, C., {Egorov}, O.~V.,
  {Garc{\'\i}a-Rodr{\'\i}guez}, A., {Glover}, S. C.~O., {Grasha}, K.,
  {Henshaw}, J.~D., {Ho}, I.~T., {Kepley}, A.~A., {Kim}, J., {Klessen}, R.~S.,
  {Kreckel}, K., {Koch}, E.~W., {Kruijssen}, J.~M.~D., {Larson}, K.~L., {Lee},
  J.~C., {Lopez}, L.~A., {Machado}, J., {Mayker}, N., {McElroy}, R., {Murphy},
  E.~J., {Ostriker}, E.~C., {Pan}, H.-A., {Pessa}, I., {Puschnig}, J., {Razza},
  A., {S{\'a}nchez-Bl{\'a}zquez}, P., {Santoro}, F., {Sardone}, A.,
  {Scheuermann}, F., {Sliwa}, K., {Sormani}, M.~C., {Stuber}, S.~K., {Thilker},
  D.~A., {Turner}, J.~A., {Utomo}, D., {Watkins}, E.~J., \& {Whitmore}, B.
  2021, \apjs, 257, 43


\bibitem[{{Li} {et~al.}(2021){Li}, {Chu}, {Raymond}, {Leibundgut},
  {Seitenzahl}, \& {Morlino}}]{cjli2021}
{Li}, C.-J., {Chu}, Y.-H., {Raymond}, J.~C., {Leibundgut}, B., {Seitenzahl},
  I.~R., \& {Morlino}, G. 2021, \apj, 923, 141


\bibitem[{{Martins} \& {Palacios}(2022)}]{martins2022}
{Martins}, F. \& {Palacios}, A. 2022, \aap, 659, A163


\bibitem[{{Martizzi} {et~al.}(2015){Martizzi}, {Faucher-Gigu{\`e}re}, \&
  {Quataert}}]{martizzi2015}
{Martizzi}, D., {Faucher-Gigu{\`e}re}, C.-A., \& {Quataert}, E. 2015, \mnras,
  450, 504


\bibitem[{{Matsuda} {et~al.}(2020){Matsuda}, {Tanaka}, {Uchida}, {Amano}, \&
  {Tsuru}}]{matsuda2020}
{Matsuda}, M., {Tanaka}, T., {Uchida}, H., {Amano}, Y., \& {Tsuru}, T.~G. 2020,
  \pasj, 72, 85


\bibitem[{{Meingast} {et~al.}(2021){Meingast}, {Alves}, \&
  {Rottensteiner}}]{meingast2021}
{Meingast}, S., {Alves}, J., \& {Rottensteiner}, A. 2021, \aap, 645, A84


\bibitem[{{Naz\'e} {et~al.}(2001){Naz\'e}, {Chu}, {Points}, {Danforth},
  {Rosado}, \& {Chen}}]{naze2001}
{Naz\'e}, Y., {Chu}, Y.-H., {Points}, S.~D., {Danforth}, C.~W., {Rosado}, M.,
  \& {Chen}, C.-H.~R. 2001, \aj, 122, 921

\bibitem[{{Norman} {et~al.}(1988){Norman}, {Dickel}, {Livio}, \&
    {Chu}}]{norman1988} {Norman}, M.~L. and {Dickel}, J.~R. and
  {Livio}, M. and {Chu}, Y.~H. 1988, in Supernova Remnants and the
  Interstellar Medium: IAU Colloquium 101, eds.\ R.~S. {Roger},
  R.~S. and T.~L.  {Landecker}. (Cambridge: Cambridge U. Press), 222.

\bibitem[{{O'Connor} \& {Ott}(2011)}]{oconnor2011}
{O'Connor}, E. \& {Ott}, C.~D. 2011, \apj, 730, 70


\bibitem[{{Oey}(1996)}]{oey1996}
{Oey}, M.~S. 1996, \apj, 467, 666


\bibitem[{{Oey} \& {Kennicutt}(1998)}]{oey1998}
{Oey}, M.~S. \& {Kennicutt}, R.~C., J. 1998, \pasa, 15, 141


\bibitem[{{Oey} {et~al.}(2009){Oey}, {Meurer}, {Yelda}, \& {Furst}}]{oey2009}
{Oey}, M.~S., {Meurer}, G.~R., {Yelda}, S., \& {Furst}, E.~J. 2009, \apss, 324,
  205


\bibitem[{{Oey} {et~al.}(2023){Oey}, {Sawant}, {Danehkar}, {Silich}, {Smith},
  {Melinder}, {Leitherer}, {Hayes}, {Jaskot}, {Calzetti}, {Chu}, {James}, \&
  {{\"O}stlin}}]{oey2023}
{Oey}, M.~S., {Sawant}, A.~N., {Danehkar}, A., {Silich}, S., {Smith}, L.~J.,
  {Melinder}, J., {Leitherer}, C., {Hayes}, M., {Jaskot}, A.~E., {Calzetti},
  D., {Chu}, Y.-H., {James}, B.~L., \& {{\"O}stlin}, G. 2023, \apjl, 958, L10


\bibitem[{Ohashi {et~al.}(2023)Ohashi, Tobin, J{\o}rgensen, Takakuwa, Sheehan,
  Aikawa, Li, Looney, Williams, Aso, Sharma, Choi), Yamato, Lee, Tomida, Yen,
  Encalada, Flores, Gavino, Kido, Han, Lin, Narayanan, Phuong,
  Santamar{\'\i}a-Miranda, Thieme, van~'t Hoff, de~Gregorio-Monsalvo, Koch,
  Kwon, Lai, Lee, Plunkett, Saigo, Hirano, Lam, \& Mori}]{eDISK}
Ohashi, N., Tobin, J.~J., J{\o}rgensen, J.~K., Takakuwa, S., Sheehan, P.,
  Aikawa, Y., Li, Z.-Y., Looney, L.~W., Williams, J.~P., Aso, Y., Sharma, R.,
  Choi), J. S.~I., Yamato, Y., Lee, J.-E., Tomida, K., Yen, H.-W., Encalada,
  F.~J., Flores, C., Gavino, S., Kido, M., Han, I., Lin, Z.-Y.~D., Narayanan,
  S., Phuong, N.~T., Santamar{\'\i}a-Miranda, A., Thieme, T.~J., van~'t Hoff,
  M. L.~R., de~Gregorio-Monsalvo, I., Koch, P.~M., Kwon, W., Lai, S.-P., Lee,
  C.~W., Plunkett, A., Saigo, K., Hirano, S., Lam, K.~H., \& Mori, S. 2023, The
  Astrophysical Journal, 951, 8
 \href{https://dx.doi.org/10.3847/1538-4357/acd384}{\urllinklabel}

\bibitem[{{Orozco-Duarte} {et~al.}(2023){Orozco-Duarte}, {Garc{\'\i}a-Segura},
  {Wofford}, \& {Toal{\'a}}}]{orozco-duarte2023}
{Orozco-Duarte}, R., {Garc{\'\i}a-Segura}, G., {Wofford}, A., \& {Toal{\'a}},
  J.~A. 2023, \mnras, 526, 5919


\bibitem[{{Pan} {et~al.}(2012){Pan}, {Ricker}, \& {Taam}}]{pan2012a}
{Pan}, K.-C., {Ricker}, P.~M., \& {Taam}, R.~E. 2012, \apj, 760, 21


\bibitem[{{Pikel'Ner}(1968)}]{pikelner1968}
{Pikel'Ner}, S.~B. 1968, \aplett, 2, 97


\bibitem[{{Pittard}(2019)}]{pittard2019}
{Pittard}, J.~M. 2019, \mnras, 488, 3376


\bibitem[{{Polak} {et~al.}(2023){Polak}, {Mac Low}, {Klessen}, {Teh},
  {Cournoyer-Cloutier}, {Andersson}, {Appel}, {Tran}, {Lewis}, {Wilhelm},
  {Portegies Zwart}, {Glover}, {Wang}, \& {McMillan}}]{polak2023}
{Polak}, B., {Mac Low}, M.-M., {Klessen}, R.~S., {Teh}, J.~W.,
  {Cournoyer-Cloutier}, C., {Andersson}, E.~P., {Appel}, S.~M., {Tran}, A.,
  {Lewis}, S.~C., {Wilhelm}, M. J.~C., {Portegies Zwart}, S., {Glover}, S.
  C.~O., {Wang}, L., \& {McMillan}, S. L.~W. 2023, arXiv e-prints,
  arXiv:2312.06509


\bibitem[{{Portegies Zwart} {et~al.}(2010){Portegies Zwart}, {McMillan}, \&
  {Gieles}}]{portegies-zwart2010}
{Portegies Zwart}, S.~F., {McMillan}, S.~L.~W., \& {Gieles}, M. 2010, \araa,
  48, 431


\bibitem[{{Ricker} \& {Taam}(2012)}]{ricker2012}
{Ricker}, P.~M. \& {Taam}, R.~E. 2012, \apj, 746, 74


\bibitem[{{Rogers} \& {Pittard}(2013)}]{rogers2013}
{Rogers}, H. \& {Pittard}, J.~M. 2013, \mnras, 431, 1337


\bibitem[{{Saken} {et~al.}(1992){Saken}, {Shull}, {Garmany}, {Nichols-Bohlin},
  \& {Fesen}}]{saken1992}
{Saken}, J.~M., {Shull}, J.~M., {Garmany}, C.~D., {Nichols-Bohlin}, J., \&
  {Fesen}, R.~A. 1992, \apj, 397, 537


\bibitem[{{Sch{\"a}fer} {et~al.}(2017){Sch{\"a}fer}, {Yang}, \&
  {Johansen}}]{schaefer2017}
{Sch{\"a}fer}, U., {Yang}, C.-C., \& {Johansen}, A. 2017, \aap, 597, A69


\bibitem[{{Segura-Cox} {et~al.}(2020){Segura-Cox}, {Schmiedeke}, {Pineda},
  {Stephens}, {Fern{\'a}ndez-L{\'o}pez}, {Looney}, {Caselli}, {Li}, {Mundy},
  {Kwon}, \& {Harris}}]{segura-cox2020}
{Segura-Cox}, D.~M., {Schmiedeke}, A., {Pineda}, J.~E., {Stephens}, I.~W.,
  {Fern{\'a}ndez-L{\'o}pez}, M., {Looney}, L.~W., {Caselli}, P., {Li}, Z.-Y.,
  {Mundy}, L.~G., {Kwon}, W., \& {Harris}, R.~J. 2020, \nat, 586, 228


\bibitem[{{Silich} \& {Tenorio-Tagle}(2018)}]{silich2018}
{Silich}, S. \& {Tenorio-Tagle}, G. 2018, \mnras, 478, 5112


\bibitem[{{Silich} {et~al.}(2004){Silich}, {Tenorio-Tagle}, \&
  {Rodr{\'\i}guez-Gonz{\'a}lez}}]{silich2004}
{Silich}, S., {Tenorio-Tagle}, G., \& {Rodr{\'\i}guez-Gonz{\'a}lez}, A. 2004,
  \apj, 610, 226


\bibitem[{{Smartt}(2015)}]{smartt2015}
{Smartt}, S.~J. 2015, \pasa, 32, e016


\bibitem[{{Smith} {et~al.}(2005){Smith}, {Edgar}, {Plucinsky}, {Wargelin},
  {Freeman}, \& {Biller}}]{smith2005}
{Smith}, R.~K., {Edgar}, R.~J., {Plucinsky}, P.~P., {Wargelin}, B.~J.,
  {Freeman}, P.~E., \& {Biller}, B.~A. 2005, \apj, 623, 225


\bibitem[{{Steigman} {et~al.}(1975){Steigman}, {Strittmatter}, \&
  {Williams}}]{steigman1975}
{Steigman}, G., {Strittmatter}, P.~A., \& {Williams}, R.~E. 1975, \apj, 198,
  575


\bibitem[{{Stephens} {et~al.}(2023){Stephens}, {Lin},
  {Fern{\'a}ndez-L{\'o}pez}, {Li}, {Looney}, {Yang}, {Harrison}, {Kataoka},
  {Carrasco-Gonzalez}, {Okuzumi}, \& {Tazaki}}]{stephens2023}
{Stephens}, I.~W., {Lin}, Z.-Y.~D., {Fern{\'a}ndez-L{\'o}pez}, M., {Li}, Z.-Y.,
  {Looney}, L.~W., {Yang}, H., {Harrison}, R., {Kataoka}, A.,
  {Carrasco-Gonzalez}, C., {Okuzumi}, S., \& {Tazaki}, R. 2023, \nat, 623, 705


\bibitem[{{Sukhbold} {et~al.}(2016){Sukhbold}, {Ertl}, {Woosley}, {Brown}, \&
  {Janka}}]{sukhbold2016}
{Sukhbold}, T., {Ertl}, T., {Woosley}, S.~E., {Brown}, J.~M., \& {Janka}, H.~T.
  2016, \apj, 821, 38


\bibitem[{{Toal\'a} \& {Arthur}(2014)}]{toala2014}
{Toal\'a}, J.~A. \& {Arthur}, S.~J. 2014, \mnras, 443, 3486


\bibitem[{{Toal\'a} \& {Arthur}(2016)}]{toala2016}
---. 2016, \mnras, 463, 4438


\bibitem[{{Toal\'a} \& {Arthur}(2018)}]{toala2018}
{Toal\'a}, J.~A., J.~A. \& {Arthur}, S.~J. 2018, Galaxies, 6, 80


\bibitem[{{Toraskar} {et~al.}(2013){Toraskar}, {Mac Low}, {Shara}, \&
  {Zurek}}]{toraskar2013}
{Toraskar}, J., {Mac Low}, M.-M., {Shara}, M.~M., \& {Zurek}, D.~R. 2013, \apj,
  768, 48


\bibitem[{{van Buren} \& {Mac Low}(1992)}]{van_buren1992}
{van Buren}, D. \& {Mac Low}, M.-M. 1992, \apj, 394, 534


\bibitem[{{Walch} {et~al.}(2015){Walch}, {Girichidis}, {Naab}, {Gatto},
  {Glover}, {W{\"u}nsch}, {Klessen}, {Clark}, {Peters}, {Derigs}, \&
  {Baczynski}}]{walch2015}
{Walch}, S., {Girichidis}, P., {Naab}, T., {Gatto}, A., {Glover}, S.~C.~O.,
  {W{\"u}nsch}, R., {Klessen}, R.~S., {Clark}, P.~C., {Peters}, T., {Derigs},
  D., \& {Baczynski}, C. 2015, \mnras, 454, 238


\bibitem[{{Weaver} {et~al.}(1977){Weaver}, {McCray}, {Castor}, {Shapiro}, \&
  {Moore}}]{weaver1977}
{Weaver}, R., {McCray}, R., {Castor}, J., {Shapiro}, P., \& {Moore}, R. 1977,
  \apj, 218, 377


\bibitem[{{W\"unsch} {et~al.}(2007){W\"unsch}, {Silich}, {Palou{\v{s}}}, \&
  {Tenorio-Tagle}}]{wuensch2007}
{W\"unsch}, R., {Silich}, S., {Palou{\v{s}}}, J., \& {Tenorio-Tagle}, G. 2007,
  \aap, 471, 579


\bibitem[{{Yang} {et~al.}(2007){Yang}, {Gruendl}, {Chu}, {Mac Low}, \&
  {Fukui}}]{yang2007}
{Yang}, C.-C., {Gruendl}, R.~A., {Chu}, Y.-H., {Mac Low}, M.-M., \& {Fukui}, Y.
  2007, \apj, 671, 374


\bibitem[{{Yang} \& {Johansen}(2014)}]{yang2014}
{Yang}, C.-C. \& {Johansen}, A. 2014, \apj, 792, 86


\bibitem[{{Zhang} \& {Chevalier}(2019)}]{zhang2019}
{Zhang}, D. \& {Chevalier}, R.~A. 2019, \mnras, 482, 1602


\bibitem[{{Zhang} {et~al.}(2014){Zhang}, {Wang}, {Ji}, {Smith}, {Foster}, \&
  {Zhou}}]{zhang2014}
{Zhang}, S., {Wang}, Q.~D., {Ji}, L., {Smith}, R.~K., {Foster}, A.~R., \&
  {Zhou}, X. 2014, \apj, 794, 61


\bibitem[{{Zhang} {et~al.}(2008){Zhang}, {Woosley}, \& {Heger}}]{zhang2008}
{Zhang}, W., {Woosley}, S.~E., \& {Heger}, A. 2008, \apj, 679, 639


\end{thebibliography}
\end{document}